# Observation of topological edge states of sound at a momentum away from the high-symmetry point

Bai-Zhan Xia[*], Sheng-Jie Zheng, Ning Chen, Ting-Ting Liu, Jun-Rui Jiao, Hong-Qing Dai, De-Jie Yu[*], Jian Liu

State Key Laboratory of Advanced Design and Manufacturing for Vehicle Body, College of Mechanical and Vehicle Engineering, Hunan University, Changsha, Hunan, 410082, China

Topologically protected one-way transportation of sound, mimicking the topological properties of the condensed matter, has received greatly attentions. Thus far, the topological phases and the topological edge states of sound are yielded in the vicinity of the Dirac cones fixed at the high symmetric points of the Brillouin zone. Here, we present a new type of the phononic topological insulator in the square lattice with position-variational Dirac cones along the high symmetric lines. The emergence of such Dirac cones, characterized by the vortex structure in a momentum space, is attributed to the unavoidable band crossing protected by the mirror symmetry. By rotating the square columns, these Dirac points are lifted and a complete band gap is induced because of the mirror-symmetry-breaking. Along the topological domain wall between the phononic crystals (PhCs) with the distinct topological phases stemming from the mirror symmetry inversion, we obtain a topological edge state for the neutral scalar sound which is absence of the intrinsic polarization and is uncoupled from an external field. Within a wide rotational range of the square column, the topological edge state in our PhCs evolves from a gapless one into a gapped one with a robust edge transport against cavities and disorders. Our excellent results are promising for the exploration of the new topological phenomena in the PhCs beyond the hexagonal lattices. Furthermore, the flexibility of the rotational square columns provides an interesting platform for the design of tunable topological acoustic devices.

[*] xiabz2013@hnu.edu.cn (Baizhan Xia)
djyu@hnu.edu.cn (Dejie Yu)

The intriguing discovery of topological insulators has stimulated a keen interest in the non-trivial topological states, characterized by the non-reciprocally spin-polarized edge transports(1-6). Compared with strongly spin-orbit-coupled electronic systems, the elusive topological properties in bosonic systems, such as photonic and phononic crystals, are still relative uncommon. The potential gradient for the electric transport is invalid for the photonic transport, thus the topological state of photon has to be excited in a more ingenious way. Magnetic fields, by breaking the time-reversal symmetry, successfully induced the photonic counterparts of the quantum Hall edge effects for the topologically protected one-way transport(7-15). At the optical frequencies, materials with strong magnetic responses are extremely absent. Even at the microwave frequencies, the practical application of the strong magnetic field is exceptionally inconvenient. Recently, the dynamic modulation(16-20) and the coupled helical mechanism(21-24), by emulating the effect of the time reversal symmetry breaking, have been introduced to activate the photonic topological states. The optically coupled resonators without breaking the time reversal symmetry have been suggested for the realization of the topologically protected optical delay lines and edge states in the photonic systems(25-32). By interfacing the bianisotropic photonic metacrystals, the spin-polarized one-way transport of the surface photons have also been successfully observed(33-44).

As the speed of sound is orders of magnitude less than the velocity of light, the phononic propagation possesses a smaller wavelength and a stronger phonon-phonon interaction. Furthermore, the slow group velocity and the high density of phonon could lead to the strong backscattering from defects and disorders. Topological condense matters feature the robust one-way edge states against backscattering. Thus, in analogy with photons, phonons can also get benefit from the topologically robust states, such as affording the unparalleled tolerances against defects over wide ranges. Unfortunately, due to the inherent longitudinal nature of the airborne acoustic polarization, it was once considered to be impossible to achieve a topological phononic insulator based on the common spin-orbital coupled mechanism. The topological gauge fluxes with the circulating background(45-48) and the time-asymmetric gyroscope(49-52) have been creatively introduced in the fluid acoustic systems and the gyroscopic mechanical systems. By emulating "magnetic fields", these circulating movements of the fluid and gyroscopes have successfully realized the topological one-way edge transport of sound. Due to the dynamic instability, the transmission loss and the inherent noise, it is still a grave challenge for the practical implementation of the moving background. Recently, the topological orders of sound in acoustic networks have also been induced by the spatiotemporal modulation(53, 54) and the resonant coupling(55-57). Such topological states can be contributed to the inherent spin-1/2 fermionic character. Thus, the key physics behind a phononic counterpart of the topological insulator is to yield an additional degree of freedom so as to degenerate the Dirac conical dispersion for a spin state. At the high symmetric points of the Brillouin zones, the phononic graphene-like lattices possess the Dirac cones with two-fold degeneracies(58) and the double Dirac cones with four-fold degeneracies(59-61). When the hexagonal lattice undergoes an inversion symmetry breaking, the band structures near these Dirac degenerate cones will inverse, accompanied with the topological phase transitions and the topologically protected one-way transports(62-67). The Dirac conical dispersion, attributing to the unavoidable band crossing, is protected by the point group symmetry(68). Thus, many PhCs, beyond the typical graphene lattice, maybe also associate with the Dirac conical dispersions. For the photonic crystals composed of the square lattices of dielectric rods, the position-varying Dirac conical dispersions were emerged at the Brillouin zone boundary(13, 69, 70). By breaking the time reversal symmetry, the degenerate Dirac cone can be reopened, leading to a photonic analog of the Chern insulator and the unidirectional helical edge state(13, 69). Even though the time reversal symmetry breaking is induced by the magnetic field in a theoretical photonic model, such results stimulate us to explore the Dirac conical dispersions and the topological states of bosons (optic and sound) at the momentum away from the high-symmetry points of the Brillouin zone.

In this letter, we systematically investigate the band degeneracy in a phononic square lattice. We show that the phononic square lattice with four equivalent mirror symmetries generates four pairs of Dirac cones, two pairs locating at the M-X and M-Y lines of the Brillouin zone and the other two pairs locating at the M-Γ lines. These Dirac conical dispersions, induced by unavoidable band degeneracy during the closing of the band gap, are stable. Small variations of microscopic system parameters can only vary their positions at the high symmetric lines, but not remove them. However, the mirror symmetry breaking will lift these Dirac points. When the square columns rotate from left to right, PhCs will undergo a symmetric inversion, accompanied with the opening-closing-reopening of band gap at the Dirac cones. This band inversion induces a topological phase transition, in analogy with the spin-orbit coupling. Along the domain wall between two PhCs with opposite topological phases, two topological edge states with symmetric and antisymmetric modes are generated. Within a wide rotational range of square columns, the topologically gapless edge states in our phononic crystals can evolve from a gapless one to a gapped one. In the experimental test, we observe the robust edge transport immunizing against cavities and disorders. Our topological phononic insulator opens up a new route towards the discovery of the topological states of sound based on the Dirac conic dispersion at a momentum away from the high-symmetry point, and develops a new platform for the investigation of the novel topological phenomenon of sound and applications such as the tunable topological acoustic devices.

**Phononic topological insulator**

We consider a PhC in a square lattice consisting of rotatable square columns, illustrated in Fig. 1a. The side length of the square column is $b$ and the lattice constant is $a$. Differing from the phononic hexagonal lattices whose Dirac cones are fixed at the corners of the Brillouin zone, the three-dimensional band structure (Fig. 1b) of our PhC shows that two classes of Dirac cones with linear dispersions appear at the momentum away from the high-symmetry points. The first class of Dirac cones, marked as $N_1$ and $N_2$, respectively emerge at the M-X and M-Y lines of the Brillouin zone. According to the effective Hamiltonian and the group theory(68-70), the emergency of the Dirac point $N_1$ is attribute to the unavoidable band crossing which is intrinsically protected by the bilaterally mirror symmetry $m_{v1}$, while the Dirac point $N_2$ is originated from the bilaterally mirror symmetry $m_{v2}$. Obviously, the eigenmodes at both Dirac points are antisymmetric and symmetric about the bilateral axes (seeing Fig. 1c for $N_1$ and Fig. 1d for $N_2$). These symmetric characteristics further verify that the Dirac degenerated states at the nodes $N_1$ and $N_2$ stem from the bilaterally mirror symmetries of the PhC. The other class of the Dirac cones, marked as $N_3$ and $N_4$, emerge at the M-Γ lines of the Brillouin zone. The eigenmodes of both Dirac points are anti-symmetric and symmetric about the diagonal axes of the PhC (seeing Fig. 1e for $N_3$ and Fig. 1f for $N_4$). These results indicate that the Dirac degenerated states at the nodes $N_3$ and $N_4$ are intrinsically protected by the diagonally mirror symmetries $m_{d1}$ and $m_{d2}$. The above Dirac cones originating from the isolated band degeneracy are stable, as long as the corresponding mirror symmetries are protected. Small variations of system parameters can only shift positions of Dirac points, but not lift them. As is seen in Fig. 1g, when $b/a$ is 0.616, the Dirac cones are emerged at the Brillouin zone corners ($M$ nodes). With the decrease of $b/a$, the Dirac cones induced by the bilaterally and diagonally mirror symmetries respectively move along the M-X line and the M-$\Gamma$ line. This extraordinary phenomenon stimulates us to control the position of the Dirac conical dispersion by modulating the filling ratio of the PhC. Furthermore, at the degenerated Dirac cone, the effective mass density is zero ($\rho_m=0$)(71, 72). When $\rho_m=0$, the effective sound speed will tend to infinity ($c_m\to\infty$), as $k_m = \sqrt{B_m/\rho_m}$. When $c_m\to\infty$, the effective wave number will tend to zero ($k_m=\omega/c_m\to 0$), which can induce an extraordinary transportation of sound, such as the zero phase propagation shown in Fig. 1h.

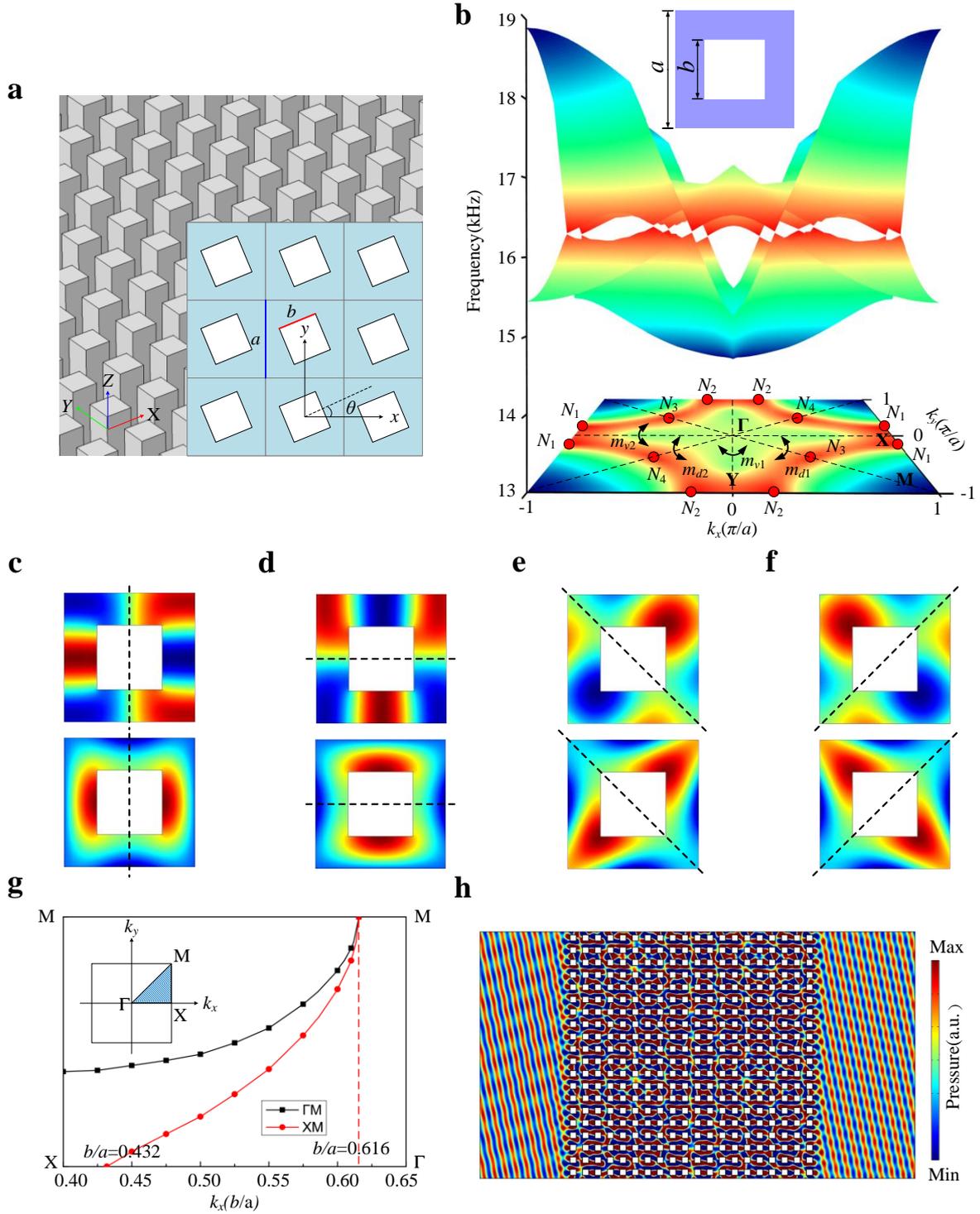

**Figure 1| a**. A geometric arrangement of PhC consisted of rotatable square columns with a lattice constant $a$=24mm. The side length of the square column is $b$=12mm. The density and the sound speed of air are 1.25kg/m$^3$ and 343m/s. The square columns are considered as rigid. **b**. The 3D band structure with Dirac cones (marked as $N_1$, $N_2$, $N_3$ and $N_4$) at the M-X/M-Y lines and the M-Γ lines of the Brillouin zone. The principal

mirror symmetric axes are $m_{v1}$, $m_{v2}$, $m_{d1}$, and $m_{d2}$. **c**. The eigenmodes at the Dirac point $N_1$ with a mirror antisymmetry and a mirror symmetry corresponding to $m_{v1}$. **d**. The eigenmodes at the Dirac point $N_2$ with a mirror antisymmetry and a mirror symmetry corresponding to $m_{v2}$. **e**. The eigenmodes at the Dirac point $N_3$ with a mirror antisymmetry and a mirror symmetry corresponding to $m_{d1}$. **f**. The eigenmodes at the Dirac point $N_4$ with a mirror antisymmetry and a mirror symmetry corresponding to $m_{d2}$. **g**. With the change of $b/a$, dependent position for the Dirac cones $N_1$ and $N_4$ respectively along the M-X line and the M-Γ line of the Brillouin zone. **h**. The profile under an incident plane wave at the Dirac frequency exhibits a zero phase modulation in the PhC.

Now, we systematically investigate the effect of the mirror symmetry breaking on these degenerated Dirac conic dispersions. By replacing the square column with a rectangular one, the bilaterally mirror symmetries ($m_{v1}$ and $m_{v2}$) of PhC are still kept, while the diagonally mirror symmetries ($m_{d1}$ and $m_{d2}$) are broken. In this case, the band structures in the vicinity of the Dirac points $N_1$ and $N_2$ are still linearly degenerated, because of its mirror symmetries about $m_{v1}$ and $m_{v2}$, while the degenerated Dirac cones $N_3$ and $N_4$ are lifted due to the mirror symmetry breaking of $m_{d1}$ and $m_{d2}$, as is seen in Fig. 2a. When the rectangular column is replaced by a triangular one with a principal axis paralleling to the mirror plane of $m_{v1}$, the Dirac cone $N_1$ is well kept. However, the other Dirac cone $N_2$ is lifted due to the mirror symmetry breaking of $m_{v2}$, shown in Fig. 2b. The eigenmodes and their symmetries of both PhCs shown in Fig. S1 and S2 in the Supplemental Material further reveal the relationship between the Dirac degenerated states of the PhC and the mirror symmetries. Note that change in the filling ratio will shift the positions of Dirac points at the M-X and M-Y lines of the Brillouin zone, but not vanishes them (Fig. 2b and 2d). Their robustness under this variation is an important signature of their topological protection. When the square column rotates from the original state with a perfect symmetry, all Dirac points ($N_1$, $N_2$, $N_3$ and $N_4$) are lifted, leading to a complete band gap, as is shown in Fig. 2e. This is due to the symmetry breaking which is stemmed from the mismatch of the mirror symmetries between the lattice and square prisms. This result is well consistent with the fact that the degenerate Dirac conical dispersion is determined by the mirror symmetry of the PhC.

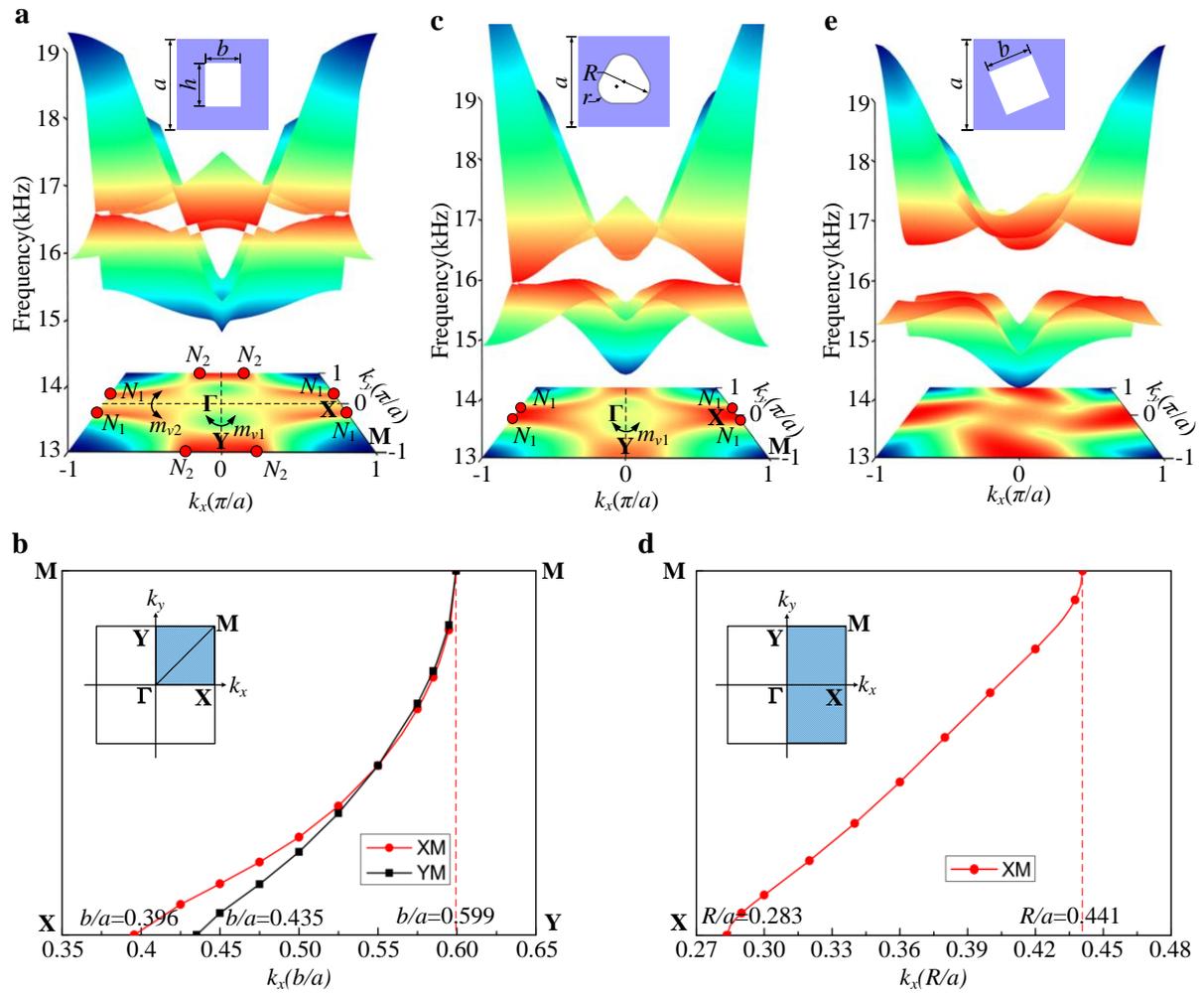

**Figure 2| a**, The 3D band structure with two pairs of Dirac cones (marked as $N_1$ and $N_2$) at the M-X and M-Y lines of the Brillouin zone. The corresponding geometric arrangement with $a$=24mm, $b$=12mm and $h$=13mm is inserted in the band structure. The principal mirror symmetries are $m_{v1}$ and $m_{v2}$. **b**, When $b/h$=12/13 does not change, with the change of $b/a$, dependent position for the Dirac cones ($N_1$ and $N_2$) along the M-X and M-Y lines of the Brillouin zone. **c**, The 3D band structure with a pair of Dirac cones (marked as $N_1$) at the M-X lines

of the Brillouin zone. The corresponding geometric arrangement with $a$=24mm, $R$=7.2mm and $r$=4.32mm is inserted in the band structure. The principal mirror symmetry is $m_{v1}$. **d**, When $R/r$=5/3 does not change, with the change of $R/a$, dependent position for the Dirac cone $N_1$ along the M-X line of the Brillouin zone. **e**, The 3D band structure without the Dirac degenerated cone for the square column with left-turn 22.5°.

**Topological edge mode**

If the degeneracy point can resemble the behavior of Weyl fermion(73), the Dirac conical dispersion will exhibit a topological nature. Here, we find that the degenerate band-edge states at the Dirac cones $N_1$ (at the M-X line) and $N_4$ (at the M-Γ line) possess two types of eigenmodes: one is the symmetric eigenmode and the other is the antisymmetric eigenmode (illustrated in Fig. 3a). If the square columns rotate apart from the specific angles $\theta=n\pi/4$ ($n$=0, ±1, ±2, …), the mirror symmetries are broken and the point group symmetry of the PhC reduces from $C_{4v}$ to $C_4$. In this case, the two-fold Dirac degeneracy, protected by the mirror symmetry at the momentum away from the high-symmetry points of the Brillouin zone, will be lifted. Angular dependent frequencies for the band-edge states near the Dirac cones $N_1$ and $N_4$ are illustrated in Figs. 3b and 3c. Evidently, the band gaps close and reopen as the square columns rotate through 0°. Noted that when the rotational angle is -45° or 45°, the mirror symmetry of the PhC is again protected, leading a new degeneration of the Dirac conical dispersion, as illustrated at Fig. 3d. The degenerated band-edge states at these new Dirac cones also possess the symmetric and antisymmetric eigenmodes (inserted in Fig. 3d). These results show that an accidental degeneracy accompanied with two band-edge states crossing each other emerges at $\theta=n\pi/4$ ($n$=0, ±1, ±2, …). Three different PC configurations, with each having the same unit cell but different rotational angles indicated by $\theta=-8^\circ$, $\theta=8^\circ$ and $\theta=37^\circ$ are considered. Their corresponding eigenmodes of the two band-edge states in the vicinity of the Dirac points $N_1$ and $N_4$ are inserted in Figs. 3b and 3c. The calculated band structures of these three PhCs are shown in Fig. S3 of the supplemental material. Phases of eigenmodes with $\theta=8^\circ$ and $\theta=37^\circ$ are approximately same. However, phases of eigenmodes with $\theta=-8^\circ$ are different from those with $\theta=8^\circ$ and $\theta=37^\circ$, exhibiting a phase inversion. Thus, during the switching of the two bands at $\theta=n\pi/4$ ($n=0, \pm1, \pm2, \ldots$), their eigenmodes will be inverted which could induce a topological phase transition. Such phenomenon can be considered as a band inversion process in the electronic systems(4-6), with $\theta=n\pi/4$ corresponding to a topological transition point in our acoustic system. Rotating columns left with $\theta=22.5^\circ$, the largest complete bandgap is yielded in the vicinity of the original Dirac cone (Fig. 2e). This complete bandgap in the bulk of the PhC induces an "insulating states". However, the phase transition, referring to the topological polarization or spin-degeneration in analogue of the electronic spin states(62-64, 67), will provide a sufficient condition for the realization of the topologically protected edge transport.

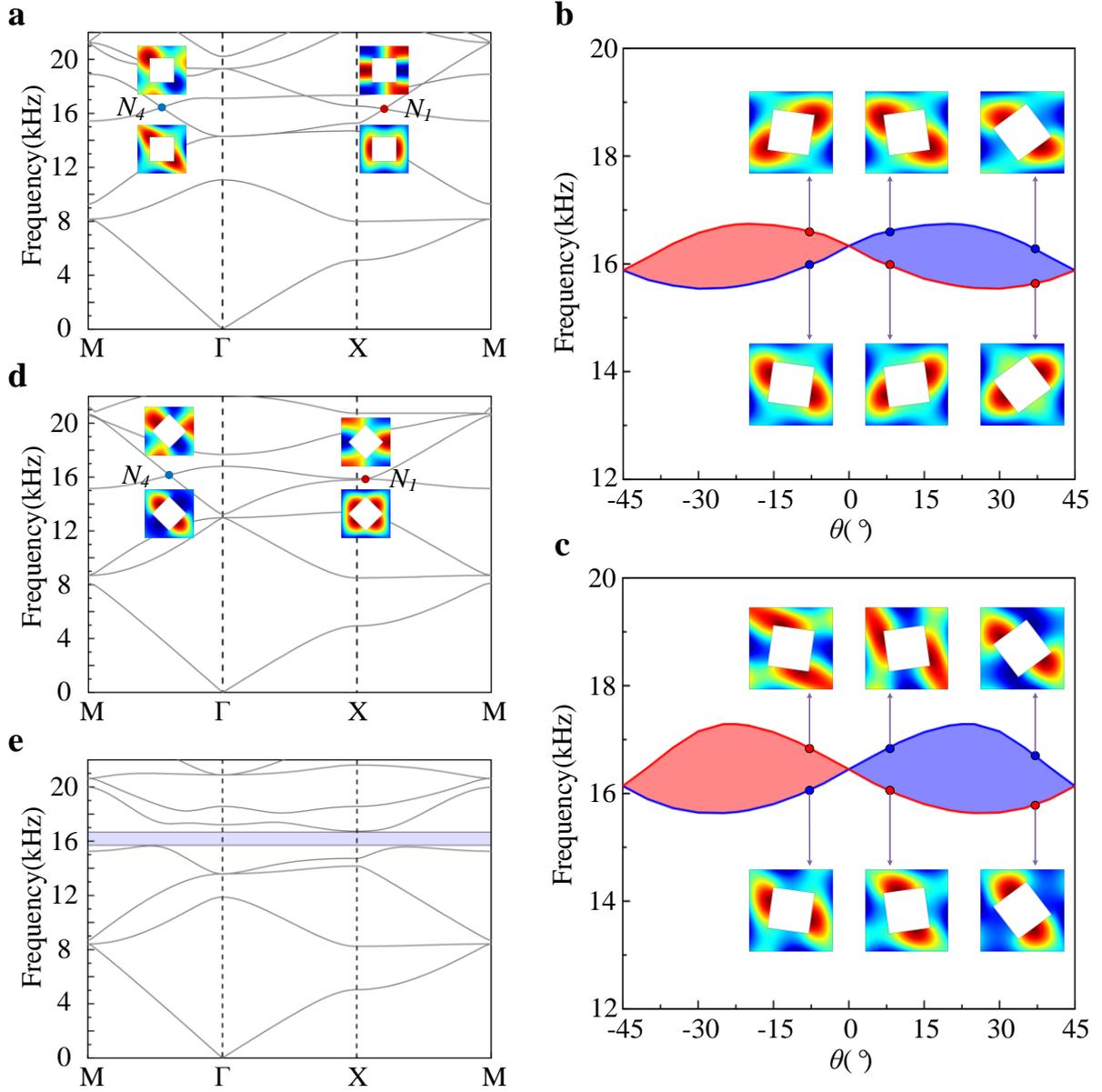

**Fig. 3, Mirror symmetry breaking and topological phase transition. a, d** Bulk band structure in the vicinity of the Dirac cone at a momentum away from the high-symmetry point at $\theta=0^o$ and $\theta=45^o$. Eigenmodes of the band-edge states are inserted in the band structure. **b, c** The angular dependent frequencies for the band-edge states in the vicinity of the Dirac cones $N_1$ and $N_4$. Eigenmodes of the band-edge states with $\theta=-8^o$, $\theta=8^o$, and $\theta=37^o$ are inserted in the band structure. **e,** Bulk band structure with a complete bandgap induced by rotating square columns with an angle $\theta=22.5^o$.

**Observation of topologically protected gapped edge states against defects**

The emergence of the topologically protected edge mode is the most fascinating property of the topological insulator. In this study, two different phononic systems are considered. The first one is a trivial phononic system with a non-topological domain wall. The second one is a nontrivial phononic system with a topological domain wall defined by the interface between reversely-rotating PhCs. For the first case, we design a representative example consisting of 20×1 square columns. 10×1 square columns locating at the upper part rotate $\theta=8^o$, while the other square columns locating the lower part rotate $\theta=37^o$. As two parts of the PhCs possess the same topological phase, a complete band gap emerges in this trivial phononic system without any edge state, shown clearly in Fig. 3a. For the second case, 10×1 square columns locating at the upper part rotate $\theta=8^o$, while the other square columns locating the lower part rotate $\theta=-8^o$. According to the bulk-boundary correspondence principle(9), the edge states can emerge at an interface separating two topological domains with opposite topological phases. The number of the edge states is determined by the number of topological phases crossing the interface(9). In this scenario, two edge bands supported by the domain wall emerge inside the overlapped bulk bandgap (shown in Fig. 3b), as the topological phases of the PhCs are reversed when crossing the domain wall. Such topological interfaces localize two edge modes formed by symmetric and anti-symmetric bonding of the evanescent waves, inserted in Fig. 3b. The extraordinary property of the topological edge state, distinguishing from the trivial counterpart, is the topological protected edge transport, analogous to the spin-polarized transport in the condensed matter systems(62-64, 67). Therefore, although the phononic bandgap, originating from the mirror symmetry breaking, efficiently blocks the sound propagation inside the bulk of the phononic system, the edge states localized at the domain walls between the distinct PhCs ensure the topologically protected edge propagation.

Furthermore, the square columns can freely rotate around their central axes, thus our phononic topological insulator can straightforward tune the topological edge domain and the corresponding topological edge state. When 10×1 square columns locating at the upper part turn from $\theta=5^o$ to $40^o$ and the others locating at the lower part correspondingly turn from $\theta=-5^o$ to $-40^o$, the bandgap and the corresponding topological edge domain can be gradually modulated. When the absolute value of the rotational angle is less than $9.25^o$ or larger than $36.48^o$, namely $\theta<9.25^o$ or $\theta>36.48^o$, the widths of the bandgap and the corresponding topological edge domain are perfectly overlapped. In this case, the topologically protected edge state is gapless (as shown in Fig. 3b). When the absolute value of the rotational angle is larger than $9.25^o$ and less than $36.48^o$, namely $9.25^o<\theta<36.48^o$, the width of the corresponding topological edge domain is less than that of the bandgap. In this case, the topologically protected edge state becomes a gapped one, not connecting the valence and conduction bulk bands. As the bandgap of the PhC for $|\theta|=22.5^o$ is widest (seeing Figs. 3b and 3c), the topological insulator consisting of square columns with an angle $\theta=-22.5^o$ and $\theta=22.5^o$ is considered. The corresponding band structure illustrated in Fig. 4d shows that the topological edge state in our phononic insulator is a gapped one which greatly differs from the gapless one in the quantum Hall effect induced by the breaking time-reversal symmetry. The topologically gapped edge state stems from the different Bloch states locating at the opposite sides of the topological domain wall. The blue and red edge states respectively express a hybridized symmetric edge mode (S) and a hybridized anti-symmetric edge mode (A) which provide a topologically protected one-way edge transport (shown in Fig. 4(b) and 4(d)). The mode profiles of the two edge states plotted in Fig. 4(b) and 4(d) also show that they have different decay lengths into the bulk from the boundary, determined largely by the rotational angles of square columns. When the rotational angle is $\theta=-22.5^o$ or $\theta=22.5^o$, the decay length is shortest, indicating the best concentration of the sound energy flow along the topological domain wall.

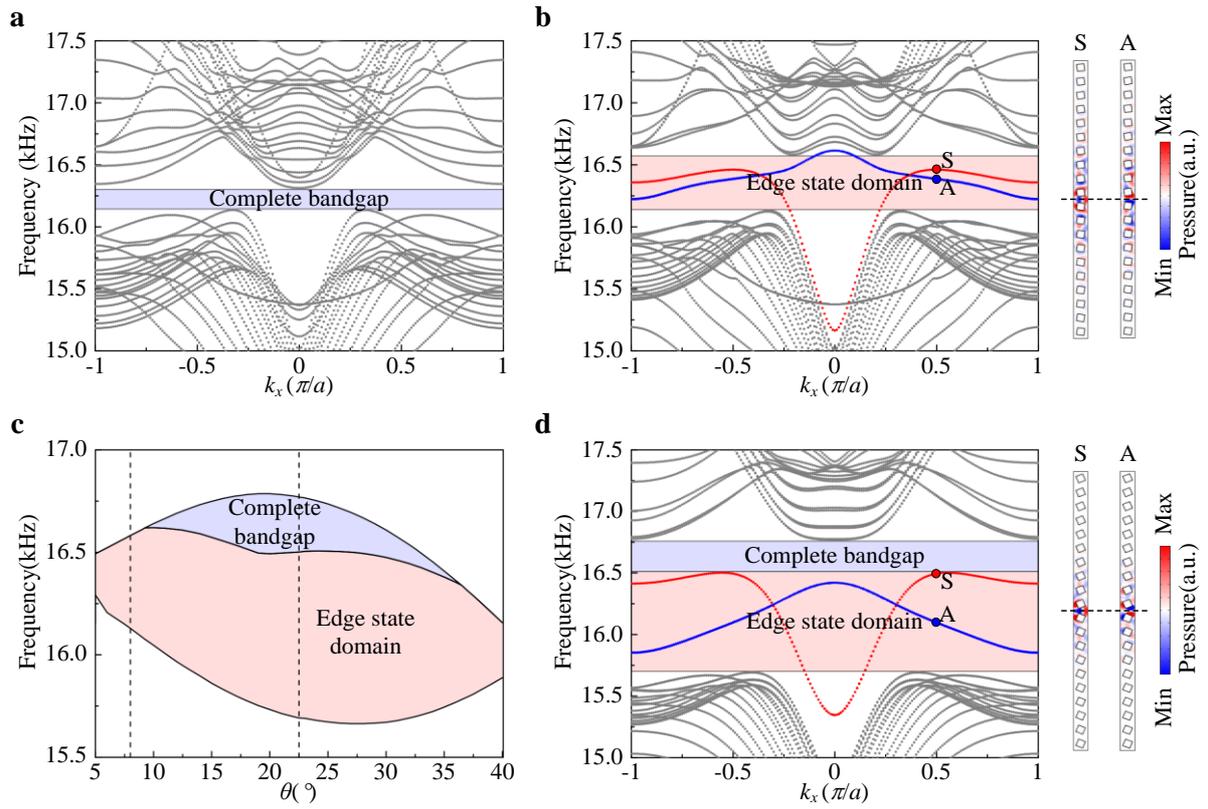

**Figure 4| Topological edge modes confined to the domain wall. a**, Phononic band structure of a supercell consisting of 20×1 square columns with a complete bandgap at the center. The square columns across the domain wall are flipped from $\theta=8^o$ to $\theta=37^o$. **b**, Phononic band structure for a supercell of 20×1 square columns with a topological edge state domain wall across which the square columns are flipped from $\theta=-8^o$ to $\theta=8^o$. **c**, The angular dependent frequencies for the bandgap and the corresponding topological edge domain. **d**, Phononic band structure for a supercell of 20×1 square columns with a nontrivial topological domain wall across which the square columns are flipped from $\theta=-22.5^o$ to $\theta=22.5^o$. The blue and red lines respectively

represent edge states confined to the domain wall. The grey lines also indicate the bulk bands. The acoustic pressure profile represents the edge modes localized at the domain wall with hybridized symmetric (S) and anti-symmetric edge modes, respectively, corresponding to the two red and blue bands at $k=0.5$ are inserted.

The topological configurations without defects and the bandgap-guiding waveguide are respectively presented in Fig. 5a and 5d. To confirm the robustness of the topological phononic edge state, we have deliberately performed large-scale simulations and experimental tests of the sound propagation along the topological domain walls with two different types of defects. The first one is an imperfect waveguide with cavities, by removing several square columns at the domain wall. Acoustic resonance arising from cavities will seriously effects the propagation property of the bandgap-guiding waveguide consisting of trivial PhCs, as is illustrated in the Fig. 5e. The second one is a disordered waveguide, by randomly rotating several square columns in arbitrary directions. For any trivial guided edge mode, the localizing and backscattering characteristics arising from strong disorders will block the sound propagation, as is illustrated in the Fig. 5f. Cavities and disorders are not the spin-mixing defects, thus the topological edge states along the domain wall will not be broken by these "nonmagnetic" impurities(9, 14, 63, 74). First-principles numerical simulations (refer to Methods for details) for the cases with two typical defects are respectively illustrated in Figs. 5b-c. These simulated acoustic pressure profiles show that the acoustic waves from the left entrance can freely detour cavities and disorders existing at the domain wall, and efficiently transport to the right exit without obviously attenuation. Thus, our topological phononic insulator can efficiently suppress the backscattering induced by the defects to a good degree, even though it does not perfectly immunize against the backscattering because of the different

Bloch states at the opposite sides of the waveguide.

A photo of our experimental set-up is in Fig.5g. Experimental transmission spectra (refer to Methods for details) measured for the topological configurations without defects and with two typical defects are plotted in Fig. 5h. The spectrum of transmission through the bulk of the PhC with all columns rotating right by an angle $\theta=22.5^o$ is also plotted in Fig. 5h. The spectrum of this PhC exhibits a very low transmission in the complete bulk bandgap from 15.7kHz to 17.0kHz. On the contrary, the phononic system with a perfect topological domain wall exhibits an extraordinarily high transmission, induced by the edge state emerging in the frequency domain 15.7kHz to 16.5kHz. Furthermore, the high transmission spectrum can be also observed for two imperfect domain walls with cavities and disorders, indicating that our phononic topological insulator can exhibit a well robust one-way transport against "nonmagnetic" defects.

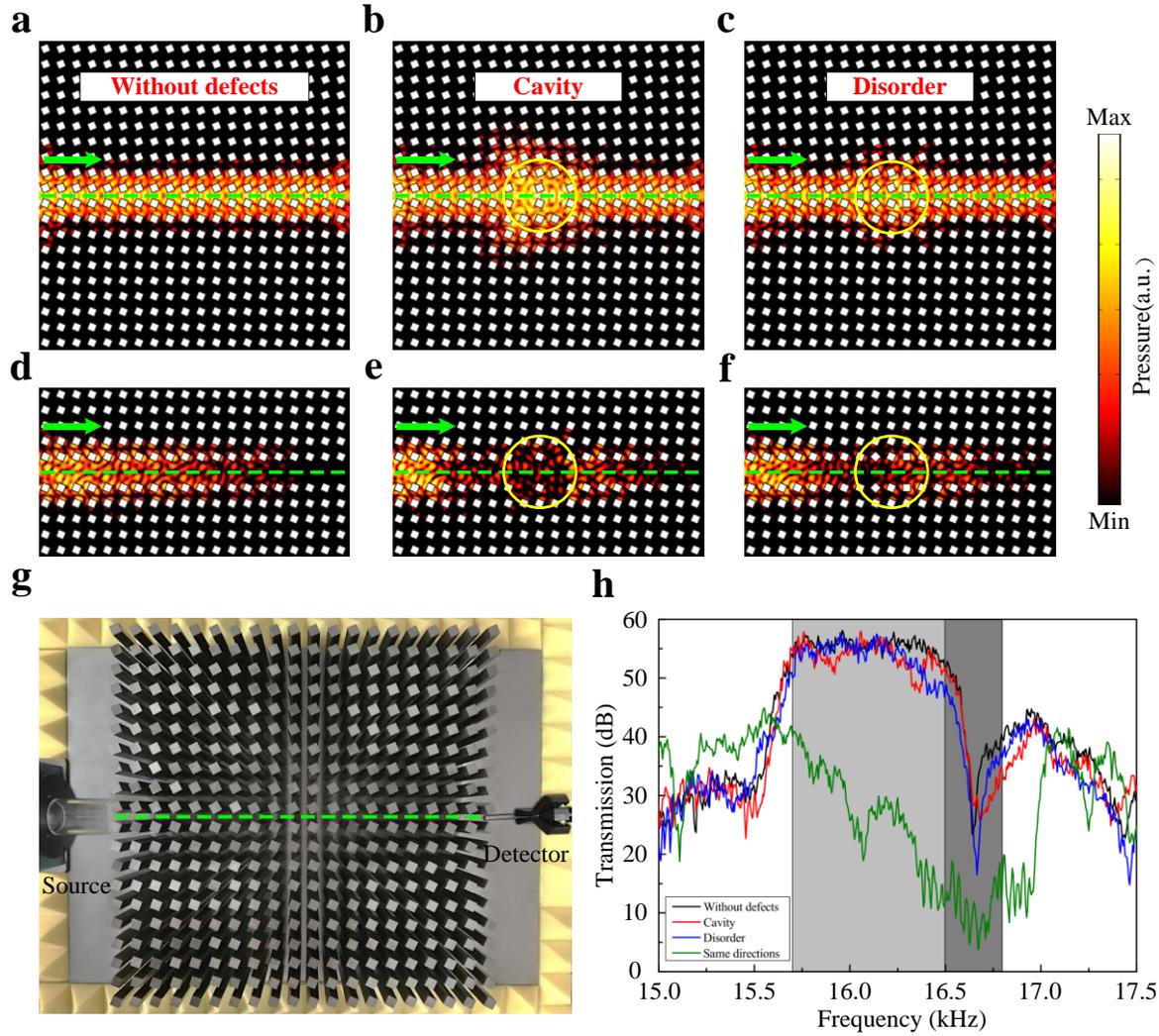

**Figure 5 | Robust one-way edge transport. a**, Simulated acoustic pressure profile in the phononic topological insulator without defects. **b-c** Simulated acoustic pressure profile in the phononic topological insulator with cavities and disorders. The frequency of the acoustic wave is 16.1 kHz (within the bulk bandgap). **d**, Simulated acoustic pressure profile in the trivial bandgap-guiding PhC waveguide without defects. **e-f**, Simulated acoustic pressure profile in the trivial bandgap-guiding PhC waveguide with cavities and disorders. **g**, A photo of our experimental set-up. The green dashed line indicates the location of PhC waveguide with $\theta=-22.5^{o}$ and $\theta=22.5^{o}$, respectively. **h**, Experimental transmission spectra of the phononic topological insulator for the acoustic wave incident from the left entrance to the right exit. The black curve corresponds to the case without defects. The red,

blue and green curves respectively indicate the transmission spectrum for the case with cavities, disorders and rotating right by an angle $\theta=22.5^{\circ}$. The shadow regions indicate the topological edge state domain.

**Summary and outlook**

To conclusion, we have shown a new type of phononic Dirac cones at the momentum away from the high-symmetry points of the square lattice. The emergence of these Dirac points is dependent on the mirror symmetry of the PhCs during the closing of bandgaps. The mirror symmetry breaking mechanism induced by the rotation of square columns can split out the Dirac cones and leads to a topological phase transition. Such an effect enables the design of topological phononic insulators with robust edge transport, even in complicated topological contours (seeing Fig. S4 with a heart-shaped bending contour and a star bending contour). The creative realization of the topological edge state in a PhC beyond the high-symmetric point can be considered as a significant development of the topological phononic insulator. Our findings also open up an intriguing avenue to explore the new topological edge states in various classical systems, from phononic systems to photonic systems, and even pave an unparalleled way for the design of novel wave functional devices.

**Acknowledgments**

The paper is supported by National Natural Science Foundation of China (No.11402083, 11572121), National Key Research and Development Program of China (2016YFD0701105) and Collaborative Innovation Center of Intelligent New Energy Vehicle and the Hunan Collaborative Innovation Center of Green Automobile.

**Author contributions**

B.-Z.X. and D.-J.Y. conceived the idea. B.-Z.X., S.-J. Z., N.C. and T.-T.L. performed the numerical simulation. S.-J. Z., H.-Q.D and J.-R.J. fabricated the samples. S.-J. Z., T.-T.L., H.-Q.D. and J.-R.J. carried out the experimental measurements. B.-Z.X., D.-J.Y., J.L., and J.M. wrote this manuscript. All the authors contributed to discussion of the results and manuscript preparation. B.-Z.X. and D.-J.Y. supervised all aspects of this work and managed this project.

**Additional information**

Supplementary information is available in the online version of the paper. Correspondence and requests for materials should be addressed to B.-Z.X., D.-J.Y. and J.L.

**Competing financial interests**

The authors declare no competing financial interests.

**Methods**

**Experiments.** The phononic crystals consist of commercial 45 steel rotatable square columns with side length $b$=12mm, arranged in air in a square lattice, the lattice constant $a$=24mm. The length tolerance of these steel square columns is ±0.1mm. Note that, in practical experiments, the square columns (of height 120 mm) are closely sandwiched between a acoustically rigid parallel plate and absorbent cottons. Experiments are conducted by a horn with a cylindrical wave tube. The sound source input is a white noise signal. The acoustic source system excites a roughly planar acoustic field, which then will be scattered at the input facet of the square phononic topological insulator according to the symmetry of the interface. In all measurements, absorbers are placed at the ends of the sample to reduce the unwanted reflection generated by the impedance mismatch between the sample and free space. The sound signal at the output facet is measured by a movable microphone (BSWA MPA421) and analyzed in LMS SCADAS III. Acoustic input frequencies are swept from 15kHz to 17.5kHz with an increment of 0.01kHz. The sample (see Fig. 5h) used to measure the dispersion is formed by two parts with identical areas, where one corresponds to the PhCs with the rotation angle $\theta=22.5°$, and the other correspond to the PhCs with $\theta=-22.5°$. Each part consists of 10×20 rods—that is, 10 1ayers along the y direction and 20 rods for each layer along the x direction. The Figs. 5g is the experimentally measured transmission.

**Simulations.** All full-wave simulations are accurately carried by a commercial FEM software (COMSOL Multiphysics). For the bulk band structures and modes of square columns (in Fig. 1b, 1c, 1d, 1e, 1f, 2e, 3a, 3b, 3c, 3d and 3e), the periodic boundary conditions are imposed along the edges of square columns to form an infinite square lattice. For bulk and edge band structures and modes of a supercell consisting of 20×1 square columns (in Fig. 4a, 4b and 4d), layer (PML) boundary conditions are imposed along the upper and lower edges of the supercell, while the periodic boundary conditions are imposed along the left and right edges of the supercell. Large-scale simulations (Figs 5a, 5b, 5c, 5d, 5eand 5f) were performed with perfectly matched layer (PML) boundary conditions around the structures. Edge modes of our phononic topological insulators are excited by a source placed at the input facets.